\documentclass[conference]{IEEEtran}
\IEEEoverridecommandlockouts
\usepackage{cite}
\usepackage{amsmath,amssymb,amsfonts}
\usepackage{algorithmic}
\usepackage{graphicx}
\usepackage{textcomp}
\usepackage{xcolor}
\usepackage{booktabs}
\usepackage{color}
\usepackage{multirow}
\usepackage{multicol}
\usepackage{adjustbox}
\usepackage[colorlinks,
            linkcolor=red,       
            anchorcolor=blue,  
            citecolor=green,
            ]{hyperref}
\usepackage{bbding}

\def\BibTeX{{\rm B\kern-.05em{\sc i\kern-.025em b}\kern-.08em
    T\kern-.1667em\lower.7ex\hbox{E}\kern-.125emX}}
\begin{document}

\title{Data-Efficient Low-Complexity Acoustic Scene Classification via Distilling and Progressive Pruning}


\author{\IEEEauthorblockN{Bing Han$^{1}$,
      Wen Huang$^{1}$,
      Zhengyang Chen$^{1}$,
      Anbai Jiang$^{2}$, 
      Pingyi Fan$^{2}$,
      Cheng Lu$^{3}$,
      Zhiqiang Lv$^{4}$, \\
      Jia Liu$^{2,4}$
      Wei-Qiang Zhang$^{2}$,
      Yanmin Qian$^{1,*}$\thanks{* Yanmin Qian is corresponding author}}
\IEEEauthorblockA{\textit{$^1$ Shanghai Jiao Tong University, Shanghai, China}}
\IEEEauthorblockA{\textit{$^2$ Tsinghua University, Beijing, China}}
\IEEEauthorblockA{\textit{$^3$ North China Electric Power University, Beijing, China}}
\IEEEauthorblockA{\textit{$^4$ Huakong AI Plus Company Limited, Beijing, China}}
\{hanbing97, yanminqian\}@sjtu.edu.cn
}

\maketitle

\begin{abstract}
The goal of the acoustic scene classification (ASC) task is to classify recordings into one of the predefined acoustic scene classes. However, in real-world scenarios, ASC systems often encounter challenges such as recording device mismatch, low-complexity constraints, and the limited availability of labeled data. To alleviate these issues, in this paper, a data-efficient and low-complexity ASC system is built with a new model architecture and better training strategies. Specifically, we firstly design a new low-complexity architecture named Rep-Mobile by integrating multi-convolution branches which can be reparameterized at inference. Compared to other models, it achieves better performance and less computational complexity. Then we apply the knowledge distillation strategy and provide a comparison of the data efficiency of the teacher model with different architectures. Finally, we propose a progressive pruning strategy, which involves pruning the model multiple times in small amounts, resulting in better performance compared to a single step pruning. Experiments are conducted on the TAU dataset. With Rep-Mobile and these training strategies, our proposed ASC system achieves the state-of-the-art (SOTA) results so far, while also winning the first place with a significant advantage over others in the DCASE2024 Challenge.

\end{abstract}


\section{Introduction}
\label{sec:intro}

Acoustic scene classification (ASC)~\cite{ASC_task} is a basic audio processing task that identifies and classifies audio signals into predefined environmental sound scenes such as airports, parks and urban streets. Due to the limitations of application scenarios, ASC systems frequently face problems such as mismatch between recording devices, constraints due to low model complexity, and a scarcity of labeled data, especially when aiming for real-time processing and deployment on resource-constrained devices~\cite{martin2022low}.

In order to build low-complexity ASC systems, many excellent works have emerged. In terms of lightweight model architecture design, researchers predominantly focus on harnessing the power of convolutional neural networks (CNNs). Conventional methods for developing CNN-based ASC systems typically rely on stacking multiple two-dimensional (2D) kernels~\cite{koutini2019receptive, hu2021two}, whereas more recent studies~\cite{cho2019acousticsceneclassificationbased,kim2022qti} have investigated the use of one-dimensional (1D) kernels as a promising alternative. In terms of training strategies, some researchers aim to enhance the robustness of models through data augmentation~\cite{morocutti2023devicerobustacousticsceneclassification,specaugpp}, while others directly incorporate adversarial strategies into training~\cite{gharib2018unsupervisedadversarialdomainadaptation}. In addition, common model compression methods including pruning~\cite{Byttebier2021} and distillation~\cite{10096110} have also been explored in ASC for further reducing computational complexity. Although these works have made good progress in building low-complexity ASC models, there has been no previous exploration of system construction methods in limited training data, and the exploration of distillation and pruning methods is also preliminary.


In this paper, to achieve the highest possible accuracy with limited parameter, computational and training data, we mainly design our system from the aspects of model design, training strategies, model pruning, and distillation on the basis of previous works. The main contributions of this paper can be summarized: 
\begin{itemize}
    \item For building low-complexity ASC model, we propose Rep-Mobile by integrating multi-branches convolution which can be reparameterized at the inference.
    \item We explore and compare the performance of teacher models with different architecture for knowledge distillation in the absence of data, and find that transformer-based models are more data-efficient.
    \item We propose progressive pruning strategy, which involves pruning the model multiple times in small amounts. It effectively reduces the parameter gap between the models before and after pruning, while resulting in better performance compared to the single step pruning method.
\end{itemize}

\section{Approaches}
In this section, we will give a detailed description on our solution for a data-efficient low-complexity ASC system, including the architecture of Rep-Mobile, knowledge distillation with ensemble teachers and progressive pruning strategy.

\subsection{Rep-Mobile Architecture}

To improve the performance of the low-complexity model for ASC task, we develop the architecture of Rep-Mobile. Like prior works on structural reparameterization~\cite{Ding_2019_ICCV,Ding_2021_CVPR,Vasu_2023_CVPR}, the train- and inference-time architecture of Rep-Mobile is different.
The blocks of Rep-Mobile are based on CP-Mobile~\cite{koutini2021receptive} which is built by enhancing MobileNet~\cite{howard2019searching} with receptive field regularization techniques ~\cite{koutini2019receptive}. Furthermore, we introduce trivial over-parameterization branches which provide further accuracy gains. 

\begin{figure}[th]
    \centering
    \includegraphics[width=1.0\linewidth]{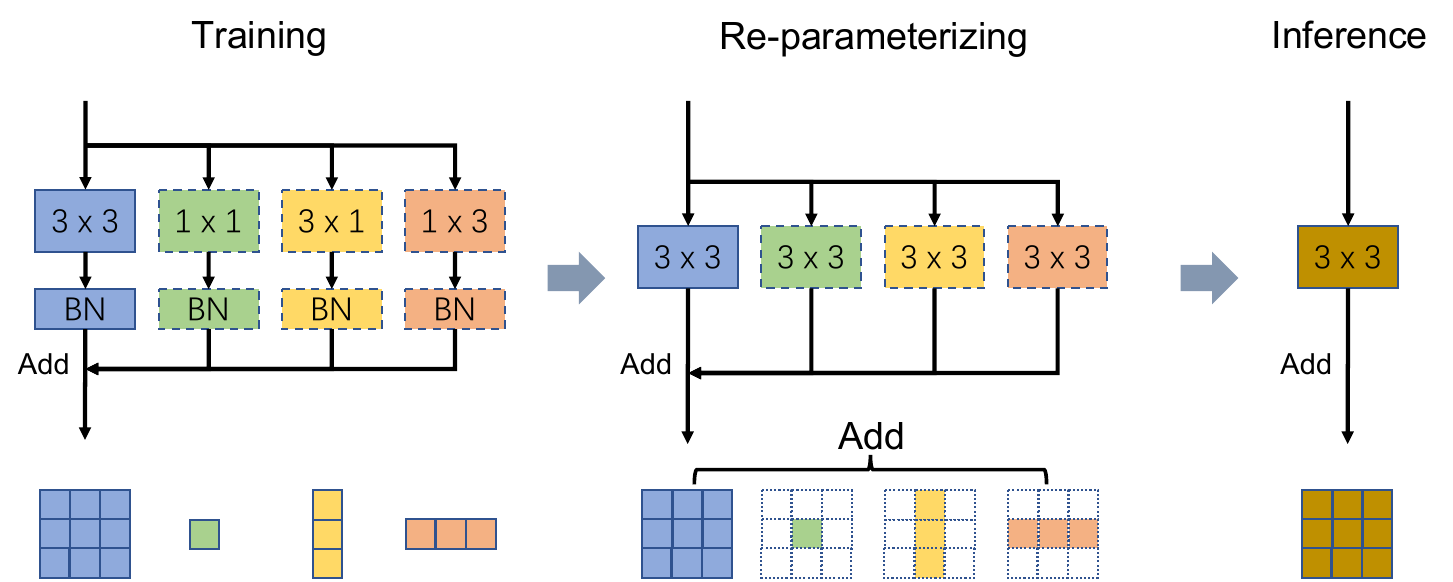}
    \caption{The reparameterization process of CNN block in Rep-Mobile. During training, multiple branches are used to enhance modeling ability, and multiple branches are merged through reparameterization without increasing computational complexity during the inference. BN denotes BatchNorm layer.}
    \label{fig:repara}
\end{figure}

The entire module design is shown in Figure~\ref{fig:repara}. During the training phase, we introduce additional branches of other shapes to the depth-wise convolution modules. The addition of a $1\times 1$ convolution kernel aims to introduce more linear transformations, while the $3\times 1$ and $1\times 3$ convolutions can extract the features separately in the frequency- and time-domains, whose similar designs have been validated as effective in other works~\cite{cai2024tf,han2022mlp}. After the training is completed, reparameterization techniques can be used to equivalently transform multiple branches by expanding the kernels with different shape, and then ultimately merge them into the main branch. Specifically, for an additional convolutional layer of kernel size $K_0\times K_1$, input channel dimension $C_{in}$ and output channel dimension $C_{out}$, the weight matrix is indicated as $\textbf{W}\in \mathbb{R}^{C_{out}\times C_{in} \times K_0 \times K_1}$ and bias is indicated as $\textbf{b} \in \mathbb{R}^D$. The first step of reparameterization is to fill in the same shape using zero values according to the shape of the main branch, denoted as $\textbf{W}'$ and $\textbf{b}'$. 
Then the next step is to perform an equivalent transformation to eliminate the BatchNorm(BN). For a BN layer, it contains accumulated mean $\mu$, standard deviation $\sigma$, scale $\gamma$ and bias $\beta$. Since convolution and BN at inference are linear operations, they can be folded by: $\hat{\textbf{W}}=\textbf{W}' \times \frac{\gamma}{\sigma}$ and $\hat{\textbf{b}}=(\textbf{b}'-\mu) \times \frac{\gamma}{\sigma} + \beta$. Following these steps, Rep-Mobile model does not have any branches which will bring additional calculation cost at inference.

\subsection{Knowledge Distillation with Ensemble Teachers}
To enhance the performance of low-complexity models in the ASC task, we employ the widely adopted knowledge distillation framework~\cite{cai2024distill}, which emphasizes replicating the final predictions of the teacher model. The knowledge transfer process consists of two key steps, beginning with the pretraining phase. At this stage, we mainly consider two different model architectures, including CNN (i.e. BC-ResNet~\cite{kim2022qti}, CP-ResNet and CP-Mobile~\cite{koutini2019receptive}) and Transformer (i.e. PaSST~\cite{koutini2021efficient}), and pretrain or finetune them on the training set. These models will be fused together to form the ensemble teacher model for subsequent knowledge distillation.

\begin{figure}[th]
    \centering
    \includegraphics[width=1.0\linewidth]{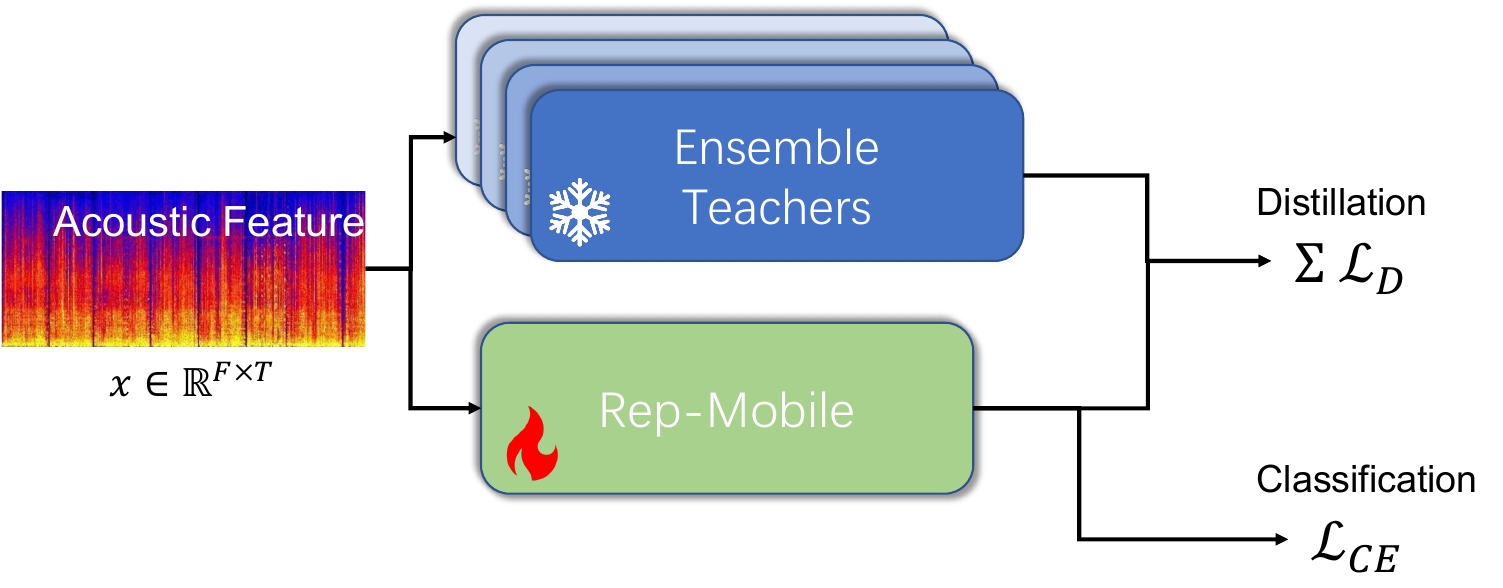}
    \caption{Illustration of knowledge distillation with ensemble teachers to low-complexity Rep-Mobile. Note that snowflake represents that the parameters of teachers are frozen, while the flame icon representes that the student model are updated with gradient.}
    \label{fig:distillation}
\end{figure}

The overview of distillation is illustrated in Figure~\ref{fig:distillation}. 
The input acoustic feature is a log-mel spectrogram $x\in \mathbb{R}^{F\times T}$. Then the features $x$ will be fed into pretrained $K$ ensemble teacher models and the corresponding predictions $y_{t,k}$ are obtained, serving as the teacher logits for the knowledge distillation process. For the student model, the acoustic feature $x$ is also used as input and $y_{s}$ is the corresponding output. Two combined losses $\mathcal{L}=\lambda \mathcal{L}_{CE} + (1-\lambda) \frac{1}{K}\sum_{k} \mathcal{L}_{D,k}$ are used for gradient optimization at the same time, where $\lambda$ is the factor to balance them.
Among them, $\mathcal{L}_{CE}$ stands for classification loss, which is calculating the cross entropy (CE) between $y_s$ and the groundtruth $\hat{y}$. $\mathcal{L}_{D,k}$ represents knowledge distillation loss, which is achieved by calculating the KL divergence between output distributions of all teacher models and students:
\begin{equation}
    \mathcal{L}_{D,k} = \textsc{KLD}(\delta(y_s), \delta(y_{t,k}))
\end{equation}
where $\delta(y) = Softmax(y/\tau)$ is a sharpening operation with temperature factor $\tau=0.1$.

In addition, to reduce the training cost, following~\cite{dinkel2024ced}, we efficiently distill the ensemble large teacher models by storing the teacher’s logits as well as their respective augmentation method on disk. As a result, during the distillation, the forward process of the teacher models does not need to be recomputed, significantly reducing computational overhead.

\subsection{Progressive Pruning}
Currently model pruning is a very commonly used compression technique~\cite{Singh2021}. It aims to reduce model complexity by removing less important parameters, such as weights or neurons, while preserving the performance. However, dropping too many parameters at once can easily lead to parameter gaps and a significant degradation in performance. 
To alleviate this issue, we propose a progressive pruning strategy which breaks down the pruning process from a single step to multiple steps. It repeats the ``pruning-finetuning'' process, and reduces the model size gradually. The goal is to create a more efficient model with fewer parameters and lower computational costs, making it ideal for resource-constrained environments like mobile or edge devices. The basic pruning algorithm we adopt is the linear pruning whose sparsity ratio increases linearly during each pruning rounds, provided by NNI toolkit~\cite{nni2021}. 


\section{Experiments Setup}
\subsection{Datasets}
The experiments are conducted with the \textit{TAU Urban Acoustic Scene 2022 Mobile} dataset~\cite{heittola2020acoustic,mesaros2018multi} following the setup in~\cite{schmid2024data}, which is a widely recognized benchmark for the task of low-complexity ASC. It provides one-second audio snippets with a sampling rate of 44.1 kHz in single-channel from ten distinct acoustic scenes. 
The recordings in this dataset were captured by various mobile devices across multiple cities worldwide, introducing challenges to the generalization ability of ASC model. 

\subsection{Training Configuration}
For the acoustic features, all audio segments are down-sampled to 32kHz. Short-Time Fourier Transform (STFT) is employed to extract time-frequency representations, with a window size of 3072 and a hop size of 512. Following the STFT, a Mel-scaled filter bank with 256 frequency bins and 4096 FFT is applied to transform the spectrograms into a Log-Mel spectrograms. To ensure a fair comparison, all models, including the baseline systems, are trained for 150 epochs using the Adam optimizer with Pytorch Framework. The batch size is fixed at 128. A warmup strategy is employed to facilitate faster convergence and stabilize training. Specifically, the learning rate is linearly increased from 0 to 0.01 in the first 5 epochs and then gradually reduced to 0 over the remaining epochs following a cosine annealing schedule. 

\subsection{Data Augmentation}
Data augmentation has a crucial impact on model performance, and we have adopted the following data augmentation strategies:
(1) Roll Audios: Waveforms are randomly rolled over time with a maximum shift of 125ms for diversity. (2) SpecAug~\cite{park2019specaugment}: Random masking in the frequency band will be applied with a maximum size of 48. (3) Freq-MixStyle~\cite{kim2022domain,schmid2023cp}:  Freq-MixStyle (FMS) is a frequency-wise version of the original MixStyle~\cite{zhou2021domain} that operates on the channel dimension. $\alpha$ and $p$ of Freq-MixStyle are respectively set to 0.3 and 0.7.


\subsection{Evaluation}
To evaluate data-efficiency, five pre-defined subsets~\cite{schmid2024data} will be utilized, each progressively reducing the available training data to 100\%, 50\%, 25\%, 10\%, and 5\% of the recordings. The distribution of acoustic scenes, cities, and recording devices remains consistent across all subsets. The smaller subsets are fully included in the larger ones, corresponding to the idea of progressively collecting more data.

\section{Results and Analysis}
\subsection{Comparison with other models}

\begin{table}[h]
    \centering
    \caption{Evaluation results on the test set under 100\% training dataset. Acc. denotes the top-1 accuracy (\%) on the test set. MACs (Multiply Accumulate Operations) indicates the computational costs per inference. Param. represents the number of parameters. Base channel of Rep-Mobile is 32.}
    \begin{tabular}{l|rr|cc}
    \toprule
       Models & \# Param. & MACs & Acc. \\ \midrule
        ResNet34~\cite{he2016deep} & 7.42M & 4901M & 58.32 \\
        ECAPA-TDNN~\cite{desplanques2020ecapa} & 6.45M & 367M & 51.28 \\
        MobileNet~\cite{seo2021mobilenet} & 3.91M & 89M & 58.22 \\ \midrule
        BC-ResNet~\cite{kim2022qti} & 23K & 29M & 55.80  \\
        CP-ResNet~\cite{koutini2021receptive} & 94K & 29M & 57.57  \\
        CP-Mobile~\cite{koutini2021receptive} & 126K & 29M & 56.15 \\
        TF-SepNet~\cite{cai2024tf} & 126K & 29M & 57.00 \\
       \midrule
       Rep-Mobile  & 126K & 29M & \textbf{58.59} \\
         \bottomrule
    \end{tabular}
    \vspace{-0.3cm}
    \label{tab:overview}
\end{table}

To demonstrate the low-complexity of our proposed model Rep-Mobile, we select multiple classic models as baseline for comparison. And results with parameters and MACs are showed in Table~\ref{tab:overview}. For a more comprehensive comparison, we firstly provide the comparison results of ResNet~\cite{he2016deep,zeinali2019but}, ECAPA-TDNN~\cite{desplanques2020ecapa}, MobileNet~\cite{seo2021mobilenet,jiang2023thuee} which are powerful audio encoders widely used in other tasks. 
Although these models achieve relatively high accuracies, the accompanying large computational and parameter loads can bring serious real-time and storage issues. 
As for the recently popular models with low-complexity design, we align the computational complexity by adjusting the width of them, in order to make more consistent comparisons. 
BC-ResNet~\cite{kim2022qti} uses 1D and 2D CNN features together for better efficiency and won $1^{st}$ in DCASE2021 challenge. It has a smaller number of parameters, but the computational complexity is relatively higher and the performance is relatively worse. CP-ResNet and CP-Mobile~\cite{koutini2021receptive} are modified CNN models by restricting the receptive field~\cite{koutini2021receptive}. And TF-SepNet~\cite{cai2024tf} separates the feature processing along the time and frequency dimensions to reduce computation cost.  Through comparison, it can be concluded that our proposed Rep-Mobile has achieved the state-of-the-art (SOTA) currently under similar computational and parameter requirements, which also proves the effectiveness of reparameterization for building low-complexity ASC models.

\subsection{Evaluation on the Multi-branch kernels}

\begin{table}[h]
    \centering
    \caption{Ablation Study of multi-branch kernels. Results are given with accuracy (\%) under 100\% training data. w/o denote without. The MACs and channels of models is 29M and 32.}
    \begin{tabular}{c|ccccc}
    \toprule
       Kernel & - & w/o 3x3 & w/o 1x1 & w/o 3x1 and 1x3 \\ \midrule
        Acc. & \textbf{58.59} & 57.06 & 57.88 & 57.47\\
         \bottomrule
    \end{tabular}
    \vspace{-0.3cm}
    \label{tab:ablation}
\end{table}

Table~\ref{tab:ablation} shows the impacts of multi-kernels within the branches in Rep-Mobile blocks. Noted that experiments are conducted under 100\% training data. When we remove any branch, the performance of the model decreases accordingly. This also indicates that convolution kernels of different shapes with prior knowledge can indeed enhance the modeling ability of the model. Moreover, benefited by reparameterization technology, this performance gain does not come with additional computational complexity.

\subsection{Evaluation on the Knowledge Distillation with Ensemble}

\begin{table}[h]
    \centering
    \caption{Evaluation Accuracy (\%) results of CNN and transformer-based teacher models with different training data sizes. Tran. denotes transformer based models. PT means pretrained on AudioSet~\cite{audioset}. Avg. is short for average. }
    \begin{tabular}{l|ccccccc}
    \toprule
       Models & 5\% & 10\% & 25\% & 50\% & 100\% & Avg.\\ \midrule
       CP-ResNet~\cite{koutini2021receptive} & 44.71 & 50.39 & 56.35 & 59.25 & 61.71 & 54.48\\
       CP-Mobile~\cite{koutini2021receptive} & 43.80 & 50.12 & 54.71 & 58.95 & 60.90 & 53.70\\
       BC-ResNet~\cite{kim2022qti} & 40.50 & 45.73 & 52.77 & 55.46 & 59.31 & 50.75 \\ 
       \midrule
       PaSST~\cite{koutini2021efficient}  & 50.90 & 53.18 & 56.98 & 58.86 & 60.20 & 56.02\\
       PaSST-PT~\cite{koutini2021efficient}  & 50.82 & 53.46 & 56.76 & 58.94 & 60.07 & 56.01 \\ \midrule
       Ensemble CNN & 49.11 & 54.72 & 59.93 & 63.56 & 66.23 & 58.71 \\
       Ensemble Tran. & 52.29 & 54.90 & 58.17 & 60.26 & 61.44 & 57.41 \\
       Ensemble All & \textbf{54.38} & \textbf{58.90} & \textbf{63.19} & \textbf{66.38} & \textbf{68.25} & \textbf{62.22} \\
    \bottomrule
    \end{tabular}
    \label{tab:teachers}
\end{table}

In order to explore the data-efficiency ability of models with different structures for teacher models in knowledge distillation, we compare the performance of common CNNs (CP-ResNet~\cite{kim2022qti}, CP-Mobile~\cite{koutini2021receptive} and BC-ResNet) and Transformers (PaSST~\cite{koutini2021efficient}) structured networks under different training data size, and the corresponding results are shown in Table~\ref{tab:teachers}. 
It should be noted that for CNN based models, 
we adjusted the width of them to maximize their performance for better teacher models.
From the results, it can be seen that both CNNs and Transformers achieve comparable results when there is sufficient data (25\%-100\%), while Transformers perform better when training data is limited (5\% and 10\%), demonstrating better data-efficiency ability. Meanwhile, we also compare whether to use AudioSet~\cite{audioset} for pre-training (PaSST vs PaSST-PT), and find that the performance gain of PaSST is not obtained by pre-training with more data, but by the architectural advantages of Transformer itself.

\begin{table}[ht]
    \centering
    \caption{Evaluation Accuracy (\%) Results of knowledge distillation with different ensemble teacher models. Avg. is short for average. Student model is Rep-Mobile with 32 channels.}
    \begin{tabular}{l|ccccccc}
    \toprule
       Teachers & 5\% & 10\% & 25\% & 50\% & 100\% & Avg.\\ \midrule
          Baseline w/o KD      & 43.41 & 46.76 & 51.22 & 54.04 & 58.59 & 50.80 \\ \midrule
       Ensemble CNN & 44.48 & 48.90 & 54.75 & 57.90 & 60.65 & 53.34 \\
       Ensemble Tran. & \textbf{49.06} & \textbf{52.65} & \textbf{57.27} & 58.65 & \textbf{61.13} & \textbf{55.75} \\
       Ensemble All & 46.83 & 51.55 & 56.94 & \textbf{59.45} & 61.12 & 55.18 \\
    \bottomrule
    \end{tabular}
    \label{tab:teachers_distillation}
    \vspace{-0.3cm}
\end{table}

In Table~\ref{tab:teachers_distillation}, we also adopted different ensemble methods to explore the effectiveness of different models when distilling knowledge to Rep-Mobile. By comparing with the baseline, using larger models for knowledge distillation on low-complexity models can effectively improve the model performance, and different teacher models exhibit different characteristics. Although the CNN ensemble teacher model has shown great superiority compared to the Transformer teacher, it has not shown corresponding advantages in knowledge distillation for the student model, especially for the setups with less training data. This also indicates that the selection of teacher models is very important in knowledge distillation, and Transformer-based systems is more appropriate in this task.

\subsection{Distillation with Progressive Pruning}

\begin{table}[h]
    \centering
    \caption{Evaluation Accuracy (\%) Results of knowledge distillation with pruning. Sgl denotes the single step pruning, and Prog denotes the progressive pruning. The number of channels represents the width of the model, which reflects the size of parameters.}
    \begin{tabular}{c|c|cccccc}
    \toprule
       Channels & Prune & 5\% & 10\% & 25\% & 50\% & 100\% & Avg.\\ \midrule
       32              & \multirow{3}{*}{-}   & 49.06 & 52.65 & 57.27 & 58.65 & 61.13 & 55.75 \\ 
       64              &    & 49.61 & 54.48 & 58.48 & 60.93 & 63.47 & 57.39 \\
       96              &    & 50.84 & 55.51 & 59.10 & 62.19 & 65.25 & 58.58 \\
       \midrule
       64$\rightarrow$32  & \multirow{2}{*}{Sgl} & 50.89 & 53.92 & 58.43 & \textbf{60.42} & 62.24 & 57.18 \\ 
       96$\rightarrow$32  & & 50.29 & 54.29 & 57.37 & 59.45 & \textbf{62.30} & 56.74 \\ \midrule
       96$\rightarrow$64$\rightarrow$32  & Prog & \textbf{51.38} & \textbf{55.26} & \textbf{58.54} & 60.07 & \textbf{62.30} & \textbf{57.51} \\ 
    \bottomrule
    \end{tabular}
    \label{tab:teachers_distillation_prun}
\end{table}

In addition, we have explored pruning strategies for building low-complexity models and showed the results in Table~\ref{tab:teachers_distillation_prun}. It should be noted that these results are obtained by distilling from Transformer to Rep-Mobile with different model size. 
From the analysis of the results, the strategy of training a larger model (64 or 96 channels) first and then pruning and compressing to smaller one (channels 32) can effectively improve the capacity of small model, by comparing with method of directly distilling to smaller one. 
Our proposed progressive pruning (channels from 96 to 64 then to 32), using multiple and small compression strategies, can effectively reduce the gap and alleviate the performance degradation caused by pruning.

\subsection{Final DCASE2024 Challenge evaluation results}

Finally, we provide a brief comparison of the top three results of DCASE2024 challenge evaluation sets in Table~\ref{tab:rank}. With our proposed Rep-Mobile, ensemble transformer distillation and progressive pruning, our system achieves the first place and outperforms other teams with a significant advantage, proving the effectiveness of our strategies.

\begin{table}[h]
    \centering
    \caption{Final Eval Scores and ranks of Task1 in DCASE2024 challenge. }
    \begin{tabular}{c|ccc}
    \toprule
       Rank & $1^{st}$(Our) & $2^{nd}$ & $3^{rd}$ \\ \midrule
       Avg. Acc. (\%) & \textbf{58.46} & 57.19 & 57.15 \\
    \bottomrule
    \end{tabular}
    \label{tab:rank}
    \vspace{-0.3cm}
\end{table}

\section{Conclusion}

In this paper, towards data-efficient low-complexity acoustic scene classification, we propose Rep-Mobile by adding multi-branch kernels and reparameteriz them during the inference process. Furthermore, we provide an analysis on the selection of teacher models for knowledge distillation and find that transformer-based models are more data-efficient and useful for knowledge distillation. Meanwhile, in order to further improve the performance of low-complexity models, we propose a progressive pruning strategy. It can effectively reduce the parameter gap between the models before and after pruning, resulting in better performance compared to the single step pruning. With these methods, we achieved the SOTA results, and won the $1^{st}$ place in task 1 of DCASE2024 challenge.

\bibliographystyle{IEEEbib}
\bibliography{refs}

\end{document}